\documentclass[letterpaper, aps, tightenlines, nofootinbib, 11pt]{article}
\pdfoutput = 1

\usepackage{setspace}
\usepackage[bookmarks = false, colorlinks = true, linkcolor = blue, citecolor = purple]{hyperref}
\usepackage{shortcuts}
\usepackage{titlesec}
\usepackage{cite}
\usepackage{tocloft}
\usepackage[vcentermath]{youngtab}
\usepackage{tensor}

\usepackage{tikz}
\usetikzlibrary{arrows,decorations.pathmorphing,backgrounds,positioning,fit,petri}

\usepackage[margin = 2.5cm]{geometry}
\begin{document}

\begin{center}

\thispagestyle{empty}

\begin{flushright}
\texttt{SU-ITP-15/16}
\end{flushright}

\vspace*{5em}

{\LARGE \bf Disorder Operators in Chern-Simons-Fermion Theories}

\vspace{1cm}

{\large \DJ or\dj e Radi\v cevi\'c}
\vspace{1em}

{\it Stanford Institute for Theoretical Physics and Department of Physics\\ Stanford University \\
Stanford, CA 94305-4060, USA}\\
\vspace{1em}
\texttt{djordje@stanford.edu}\\

\vspace{0.08\textheight}
\begin{abstract}
Building on the recent progress in solving Chern-Simons-matter theories in the planar limit, we compute the scaling dimensions of a large class of disorder (``monopole'') operators in $U(N)_k$ Chern-Simons-fermion theories at all 't Hooft couplings. We find that the lowest-dimension operator of this sort has dimension $\frac23 k^{3/2}$. We comment on the implications of these results to analyzing maps of fermionic disorder operators under 3D bosonization.
\end{abstract}
\end{center}

\newpage

%\tableofcontents

\section{Introduction}

Gauge theories in three spacetime dimensions typically possess solitonic disorder operators that strongly affect the physics at long distances. A classic example is Polyakov's proof that monopoles in a pure $U(1)$ gauge theory lead to confinement even at weak coupling \cite{Polyakov:1975}.\footnote{We define these disorder operators as instantons in three dimensions or monopole states in four dimensions --- they are points in $\R^3$ that emit magnetic flux. We will give a more careful definition in Section 2, where we will also explain what we mean by monopole states in three dimensions. For a general review of basic properties of monopoles and their relation to other solitons in various dimensions, see e.g.~\cite{Tong:2005un, Preskill:1984gd}.} Effects of gauge theory disorder operators on the phase structure of many other three-dimensional models have been analyzed in e.g.~\cite{Polyakov:1976fu, Hermele:2009pw, Baskaran:1987my, Read:1989zz, Read:1990zza, Senthil:2004}. The primary question concerning these operators is whether they cause confinement; if not, as is often the case, the next order of business is to understand the IR fixed point. Identifying disorder operators in such a nontrivial CFT is often a daunting task. Using various classical limits, their quantum numbers have been calculated in the IR of QED with many fermion flavors \cite{Pufu:2013vpa, Borokhov:2002ib}, in the critical $\C\P^{N - 1}$ model at large $N$ \cite{Murthy:1989ps,Metlitski:2008dw, Dyer:2015zha}, in the IR of QCD with many fermion flavors and with various gauge groups \cite{Dyer:2013fja}, and in the infinite-level limit of conformal Chern-Simons-matter theories \cite{Kim:2009ia, Kim:2010ac}. Disorder operators also feature prominently in supersymmetric theories, where their dimensions and intricate duality mappings can often be understood using holomorphy and localization \cite{Kim:2009wb, Kim:2012uz,  Kapustin:1999ha, Borokhov:2002cg, Aharony:1997bx, Aharony:1997gp, Aharony:2013dha, Aharony:2014uya, Aharony:2015pla, Bashkirov:2011vy, Benini:2011mf, Giveon:2008zn, Intriligator:1995id, Intriligator:2013lca, Strassler:1999hy}.

Understanding disorder operators in the IR of non-Abelian theories without supersymmetry is generally difficult. Even in the planar limit, no results at finite 't Hooft coupling were available unless the number of matter flavors was much larger than the number of colors. However, recent progress in studying the 't Hooft limit of Chern-Simons (CS) theories coupled to matter \cite{Gur-Ari:2015pca, Aharony:2011jz, Aharony:2012nh, Aharony:2012ns, Anninos:2014hia,  Dandekar:2014era, Giombi:2011kc, GurAri:2012is, Inbasekar:2015tsa, Jain:2012qi, Jain:2013gza, Jain:2013py, Jain:2014nza, Minwalla:2015sca, Radicevic:2012in, Shenker:2011zf, Takimi:2013zca, Yokoyama:2012fa} makes it possible to elegantly compute the scaling dimensions of an interesting class of disorder operators in conformal CS-fermion theories with only one fermion flavor. We perform this computation in this brief note. Our main result is that the lowest-dimension disorder operators with $q$ units of magnetic flux have dimension
\bel{\label{result}
  \Delta = \frac23 \left(qk\right)^{3/2},
}
where $k$ is the Yang-Mills-regulated CS level. This result is valid at all 't Hooft couplings.

Our presentation starts with a technical definition of disorder operators in Section 2.  In Section 3 we obtain the advertised result \eqref{result} by computing the thermal partition function of all CS-fermion states with a given magnetic flux on a two-sphere, and along the way we clarify certain aspects of the path integral computation given in \cite{Jain:2013py}. In Section 4 we conclude by discussing implications of our findings for the bosonization duality between CS-fermion and CS-boson theories.

\section{Disorder operators in three-dimensional gauge theories}

We are interested in local operators $\O(x)$ that create magnetic flux in a theory with gauge group $G$. A state created by a such a disorder operator at $x \in \R^3$ has flux that can be gauge-fixed into the form
\bel{\label{def flux}
  \int_{S^2(x)} \avg F = 4\pi q_i H^i,
}
where $S^2(x)$ is any two-sphere enclosing $x$, $F$ is the gauge field strength in the theory, $H^i$ are the anti-Hermitian Cartans of $G$, and $q_i$ are the gauge-invariant GNO charges that index different monopole states in four spacetime dimensions \cite{Goddard:1976qe}.

The quantization of GNO charges is most easily demonstrated in the path integral language, where configurations with a disorder operator at the origin can be depicted (in a suitable gauge) as fluctuations around Wu-Yang gauge field configurations \cite{Wu:1975es}
\bel{\label{def WY}
  \A(r, \theta, \phi) = \frac{q_i H^i}{r} \left\{
        \begin{array}{ll}
          (1 - \cos\theta) \d \phi , & \hbox{$\theta \leq \frac\pi2$,} \\
          (-1 - \cos\theta) \d \phi , & \hbox{$\theta \geq \frac\pi2$.}
        \end{array}
      \right.
}
(Note that the field strength associated to the above connection gives $4\pi q_i H^i$ when integrated over an $S^2$ centered at the origin.) This is an allowed configuration if the gauge fields on the south and north hemispheres differ by a gauge transformation. Fixing $r = 1$ for convenience, the difference between north and south at $\theta = \pi/2$ is $2q_i H^i \d \phi$, and so the appropriate gauge transformation is enacted by the group element $e^{2 q_i H^i \phi}$. This object must be well-defined as we circle the equator of the $S^2$ by letting $\phi \mapsto \phi + 2\pi$. This condition, $e^{4\pi q_i H^i} = 1$, is only satisfied for a discrete set of $q_i$'s.

There are reasons to believe that not all GNO charges correspond to different disorder operators. First, even if the Wu-Yang backgrounds are classical saddles, they need not all be stable in the quantum theory \cite{Dyer:2013fja}. Second, the operators defined via eq.~\eqref{def flux} need not be eigenstates of the dilatation operator in the IR CFT, and so they might not all be independent. Third, the GNO monopoles are not all topologically protected. Topological charges exist only if the gauge group has a nontrivial fundamental group $\pi_1(G)$, e.g.~for $G = U(N)$ but not for $G = SU(N)$ \cite{Preskill:1984gd}. However, even though the GNO classification may be too refined, we will find disorder operators with small amounts of flux in the planar limit that appear to have well-defined scaling dimensions with no indication of instabilities. This means that the situation is similar to the one found in many-flavor QCD \cite{Dyer:2013fja}, where large classes of GNO charges were shown to correspond to stable monopoles even though there was no a priori reason for them to do so.

So far we did not assume that the theory was conformal, and the operators $\O(x)$ were instantonic in character. These instantons represent transitions to/from states with magnetic flux that we will call monopoles. Analyzing the spectrum of such states in general QFTs is hard, but progress was made in the supersymmetric context in \cite{Intriligator:2013lca}. Things are simpler in a conformal theory, where each disorder operator corresponds to a monopole state $\qvec\O$ on $S^2_R$, a two-sphere of radius $R$. In the path integral these states correspond to configurations with background fields $\A(R, \theta, \phi)$ on $S_R^2$.

In a Yang-Mills theory coupled to matter, disorder operators are defined in a gauge-invariant way using eq.~\eqref{def flux}. Their scaling dimensions are determined, via the state-operator correspondence, as Casimir energies of the fluctuations around Wu-Yang monopoles on $\R \times S^2_R$, when these are good saddles.\footnote{Wu-Yang backgrounds give rise to a constant magnetic field on the sphere, since $F$ is given by a constant times the area form. If there is a reason to break the rotational invariance in the theory, it may happen that the true saddles are large deformations away from eq.~\eqref{def WY} for which the magnetic flux is not uniformly distributed across the sphere. In that case, all comments we make apply to fluctuations around these new backgrounds. Spherical symmetry will never be broken in our examples, so we will work with fluctuations around uniform flux backgrounds.} The fluctuations in question must be gauge-invariant themselves. In the case of Yang-Mills theory this means that no operators with electric charge may be present.

This picture changes if a Chern-Simons term at level $k$ is present in the action.\footnote{We will discuss the regularization dependence of the level in the next section.} The Wu-Yang configuration in this case has electric charge $k q_i H^i$, and the Gauss law requires that matter fields compensate for this charge \cite{Aharony:2015pla, Kim:2009ia, Borokhov:2002cg, Kim:2010ac}. This means that gauge-invariant disorder operators in CS-matter theories do not merely create magnetic flux; they also create enough matter to dress themselves into an electric singlet. This dressing may introduce large corrections to the Casimir energy due to matter self-interactions \cite{Aharony:2015pla}. For instance, in a CS-fermion theory at $k \rar \infty$ (or at zero 't Hooft coupling in the planar limit), the matter fields can be heuristically treated as noninteracting fermions in a background magnetic field that fill up a Fermi sea until they reach $\sim qk$ units of charge; the Landau levels are separated by gaps of order $\sqrt{qk}$, giving a total Fermi energy of $\sim (qk)^{3/2}$ \cite{Pufu:unp}. (See also \cite{Hellerman:2015nra}.) We will show that this holds at all 't Hooft couplings.

Our approach to calculating disorder operator dimensions is the following. We focus on the conformal CS-fermion theory, radially quantized with a Wu-Yang background \eqref{def WY} on the spatial two-sphere, and we compute the path integral over all gauge field and matter fluctuations around this background. If the Euclidean spacetime is taken to be $S_\beta^1 \times S^2_R$, the result is a thermal partition function of all monopole states with the given set of GNO charges (with matter dressing automatically taken care of). In the low temperature ($\beta/R \rar \infty$) limit, the result must take the form $Z = e^{-\beta \Delta}$, with $\Delta$ being the desired energy (or flat-space scaling dimension) of the lightest disorder operator in this GNO class. We will calculate this $\Delta$ in the 't Hooft limit by using several subtle tricks that we carefully explain in the following section.

\section{Monopole states in Chern-Simons-fermion theory}

\subsection{Chern-Simons-matter in the 't Hooft limit}

Consider the conformal $U(N)_k$ Chern-Simons theory coupled to Dirac fermions in the fundamental representation. The thermal partition function is given by the path integral on $S^1_\beta \times S^2_R$, with the two-sphere volume $V_2 = 4\pi R^2$ and temperature $T = 1/\beta$. The Euclidean action is
\bel{\label{def CS}
  S = \frac{ik}{4\pi} \int \d^3 x\, \epsilon^{\mu\nu\rho}\, \Tr\left(A_\mu \del_\nu A_\rho + \frac23 A_\mu A_\nu A_\rho \right) + \int \d^3 x\ \bar\psi_i \gamma^\mu D_\mu \psi_i, \quad D_\mu = \del_\mu + A_\mu.
}
We normalize $\Tr(T^a T^b) = - \frac12 \delta^{ab}$, and in particular we choose $H^j$ to be a matrix whose only nonzero entry is $\frac i{\sqrt 2}$ at the $j$'th place on the diagonal. (This choice means that GNO charges have values $\frac{q_i}{\sqrt 2}$ for $q_i \in \Z$.) We are interested in the planar limit, where $N$ and $k$ are taken to infinity with the 't Hooft coupling $\lambda = N/k$ fixed. In dimensional regularization, the theory is unitary only for $|\lambda| \leq 1$ \cite{Aharony:2011jz, Giombi:2011kc, Aharony:2012nh}. The CS level $k$ is quantized by demanding invariance under large gauge transformations; in order to offset the parity anomaly of the fermions, $k - 1/2$ should take on values in $\Z$ \cite{Borokhov:2002ib}, but in the 't Hooft limit this is immaterial and we may think of $k$ as being a (large) integer. At $\lambda \ll 1$ we recover the singlet fermion model \cite{Shenker:2011zf}, and as $|\lambda|$ approaches unity the theory approaches pure CS theory \cite{Anninos:2014hia}.

A subtlety (pointed out by O.~Aharony) arises here because we regulate the above theory using dimensional reduction, but the preceding discussion of disorder operators in CS theory assumed regularization using a Yang-Mills term. The CS levels in the two regularizations are related by $k\_{YM} = k - N \sgn k$. In dimensional reduction this shift arises because we must regulate the monopoles by giving them a core of nonzero size. This UV regularization effectively acts as an edge of the system, and the electric charge of the monopoles gets shifted by the $N$ edge states of CS theory.

The operator content is simple to describe: CS has no dynamical degrees of freedom and the only operators with $O(N^0)$ energies are products of $U(N)$ ``single-trace'' singlets built out of two matter fields and covariant derivatives. These single-trace operators are the scalar operator $\bar \psi \psi$ and conserved currents $\bar \psi (D_\mu)^\# \psi$; they have protected ($\lambda$-independent) scaling dimensions in the planar limit, and they are conjectured to be holographically dual to higher-spin fields of the Vasiliev system in AdS$_4$ \cite{Anninos:2014hia, Giombi:2011kc, Aharony:2012nh, Shenker:2011zf}. If the gauge group is $SU(N)$, the theory also contains baryons $B = \epsilon^{i_1\ldots i_N} \psi_{i_1}\ldots \psi_{i_N}$ whose energies are $O(N)$ \cite{Shenker:2011zf}. For any gauge group, there are also disorder operators defined in Section 2, and if the group is not simply connected (e.g.~for $U(N)$) these operators may have nontrivial topological charge.

\begin{figure}[tb!]
\begin{center}

\begin{tikzpicture}[scale = 1.5]
  \draw[very thick] (-2, 2) -- (-2, -2) -- (2, -2) --  (2, 2);
  \draw[thick, blue, dashed] (2, -2) .. controls (0.5, -1.95) and (-2, -1) .. (-1.95, 2);
  \draw[thick, red, dashdotted] (-2, -1) .. controls (-1, -1) and (2, 0.5).. (1.95, 2);

  \draw (0, -2.05) node[anchor = north] {$\lambda = \frac Nk$};
  \draw (-2, -2.05) node[anchor = north] {$\lambda = 0$};
  \draw (2, -2.05) node[anchor = north] {$\lambda = 1$};
  \draw (-2.05, 0) node[anchor = east] {$\zeta = \frac{T^2}N$};
  \draw (-2.05, -1.8) node[anchor = east] {$\zeta = 0$};
  \draw (-2.05, 2) node[anchor = east] {$\zeta = \infty$};
\end{tikzpicture}

\end{center}
\caption{\small Phases of CS-matter theories in the planar limit \cite{Jain:2013py}.  The red/dash-dotted line corresponds to the Gross-Witten-Wadia (GWW) phase transition \cite{Wadia:2012fr, Gross:1980he, Wadia:1980cp}. It intercepts $\lambda = 0$ at $\zeta = O(1)$ and is the only transition in the singlet vector model \cite{Shenker:2011zf}. The blue/dashed line corresponds to the Douglas-Kazakov (DK) phase transition \cite{Douglas:1993iia}. At $\lambda = 1$ there are no transitions at any $T$, in line with the assumption that this is a pure CS theory.  }
\label{fig phases}
\end{figure}
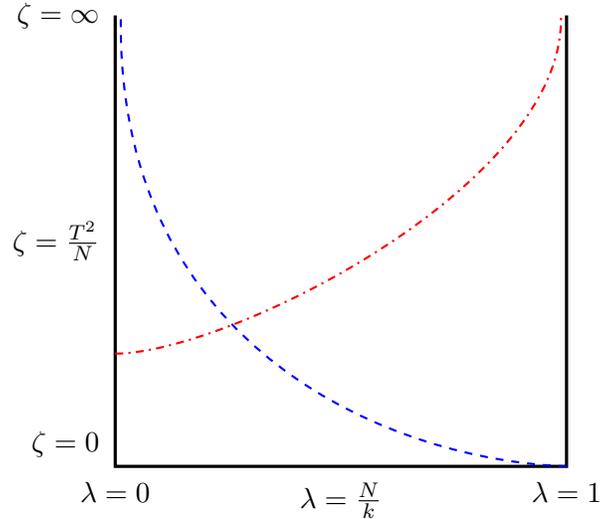

The phase structure of planar CS-fermion theory at finite temperature is very rich (see Fig.~\ref{fig phases}). At any finite 't Hooft coupling $0 < \lambda < 1$ there are two phase transitions (which overlap at a critical $\lambda$). Both transitions happen at temperatures of order $\sqrt N$. These different phases are observed in the density of eigenvalues of the matrix model obtained by integrating out matter fields from \eqref{def CS} in the high temperature limit $V_2 T^2 \sim N \gg 1$. Interpreting the transitions in terms of the original variables is still an open question, but Ref.~\cite{Shenker:2011zf} has shown suggestive evidence that baryon-antibaryon pairs introduce relations between singlet states at high energies and thereby cause a qualitative change in the density of states --- an effective ``deconfinement'' that allows one set of singlets to recombine its constituents into a different set of singlets. This transition is accompanied by going from a singlet free energy $F \sim \left(V_2 T^2\right)^2$ to a deconfined free energy $F \sim N V_2 T^2$. It has also been speculated that the second phase transition is related to the proliferation of monopoles \cite{Jain:2013py}.

A fascinating property of this planar CS-fermion theory is that it appears dual to the bosonic theory of CS gauge fields coupled to a Wilson-Fisher fixed point \cite{Minwalla:2015sca, Inbasekar:2015tsa, Dandekar:2014era, Gur-Ari:2015pca, Takimi:2013zca,Jain:2013gza, Aharony:2012ns, Jain:2014nza, Jain:2013py, Aharony:2012nh,  Anninos:2014hia}. (Similarly, the CS-Gross-Neveu model is related to the CS theory coupled to the ordinary free scalar; our calculation holds for monopoles in this other CS-fermion theory, as well.) This ``bosonization'' duality relates bosonic and fermionic theories with groups $U(N_B)_k$ and $U(N_F)_{-k}$ and with couplings that satisfy $|\lambda_F| + |\lambda_B| = 1$. This is a strong-weak duality that can be viewed both as a nonsupersymmetric Giveon-Kutasov duality \cite{Gur-Ari:2015pca, Jain:2013gza} and as a level-rank duality of pure CS theory extended to theories with matter \cite{Jain:2013py}. The bosonization conjecture holds at the level of matching partition functions, correlators, S-matrices, and single-trace operators, but it is not known how nonperturbative operators map across the duality. Since our result gives us dimensions of disorder operators at strong coupling, we will be able to tie this to the known facts about nonperturbative operators at weak coupling, and we will conjecture two possible patterns for how these nonperturbative operators may map to each other.

Let us now review the methodology of solving the CS-fermion theory in the planar limit and without monopole backgrounds. We present a streamlined version of the procedure that has been explained in great detail in Ref.~\cite{Jain:2013py}. For convenience, we will set the radius of the two-sphere to be $R = 1$; this turns the temperature $T$ into a tunable dimensionless parameter. At finite temperature, matter fields develop thermal masses proportional to $T$, so at $T \gg 1$ the fermions become very massive, propagating only on scales $\sim 1/T$ much smaller than the size of the spatial $S^2$. As for the gauge fields, the only gauge-invariant degrees of freedom are contained in the Polyakov loop, i.e.~in the holonomy of the gauge field around the thermal circle. This object is defined as the path-ordered exponential
\bel{\label{def U}
  U(\b x) = \exp\left\{\oint_{S^1_\beta} \d\tau A_3(\tau, \b x) \right\}.
}
The Polyakov loop can be viewed as a map from $S^2$ to $U(N)$. As such, it is not gauge-invariant --- it can still change under time-independent gauge transformations. Its eigenvalues are gauge-invariant, however, and below we will describe the gauge fixing needed to reduce $U(\b x)$ to its abelianized (diagonal) form $e^{\alpha_i(\b x) H^i \sqrt 2}$ with $\alpha_i \in [0, 2\pi)$.

The CS action produces a $\delta$-function potential that sharply suppresses spatial variations of the Polyakov loop eigenvalues, and the large thermal mass of the fermions ensures that integrating out matter will not change this potential on length scales greater than $1/T$. Thus we may separate the fields in \eqref{def CS} into ``high-energy'' modes (all matter fields and the large-momentum gauge fields that glue them into singlets) and ``low-energy'' modes (the Polyakov loop and the small-momentum gauge fields).

The high-energy degrees of freedom are all fields at momenta above $T$ (in particular, this includes all matter fields). Since the curvature of $S^2$ is much smaller than $T$, these fields should be thought of as living in a flat space with a constant Polyakov loop in the background \cite{Aharony:2012ns}. The gauge fields can be dealt with by fixing the light-cone gauge, $A_- \equiv (A_1 + iA_2)/\sqrt2 = 0$ \cite{Giombi:2011kc}. This choice gives a trivial Faddeev-Popov determinant. Integrating out the remaining gauge fields $A_+$ and $A_3$ gives a nonlocal potential for the matter fields, and matter can in turn be integrated out by the usual large-$N$ procedure for $U(N)$ models \cite{Moshe:2003xn}. After all the high-energy modes are integrated out, we are left with an effective potential for the Polyakov loop that takes the form of a derivative expansion:
\bel{
  v[U] = T^2 \int_{S^2} \d^2\b x \sqrt{g(\b x)} \left(v_0\big(U(\b x)\big) + \frac1{T^2} v_2\big(U(\b x), \nabla U(\b x)\big) + \ldots  \right).
}
The potentials $v_n(U)$ above are all of order $N$ and are formed by traces of arbitrarily many powers of $U(\b x)$ and $n$ powers of $\nabla U(\b x)$. At large $T$ the derivative pieces can be ignored and we can replace the Polyakov loop $U(\b x)$ with a constant matrix $U$ governed by the eigenvalue potential
\bel{\label{def v(U)}
  v[U] = 4\pi T^2 v_0(U).
}
Exact forms of this potential for various theories at all $\lambda$ have been given in Ref.~\cite{Jain:2013py}.

Integrating low-energy modes now proceeds by fixing the maximal torus gauge \cite{Blau:1993tv}. This is done in three steps \cite{Jain:2013py}:
\begin{enumerate}
  \item Fix the temporal gauge $\del_3 A_3 = 0$;
  \item Use time-independent gauge transformations to abelianize the Polyakov loop and impose $U(\b x) = e^{\alpha_i(\b x) H^i \sqrt 2}$ at each $\b x$ on the two-sphere;
  \item Use the remaining time-independent transformations to impose the Coulomb gauge for the Cartan components of spatial gauge fields $A^i_{1/2}$.
\end{enumerate}
Fixing this gauge and integrating all modes other than $U(\b x)$ (with $T$ as the UV regulator) results in a matrix model for the Polyakov loop. As we have mentioned, the gauge-fixed integral will impose a $\delta$-function on non-constant abelianized configurations $U(\b x)$, so once we go to the maximal torus gauge we may just talk about a constant matrix $U = e^{\alpha_i H^i\sqrt 2}$. The Faddeev-Popov determinants and the integration over the remaining modes (the off-diagonal spatial components of gauge fields) combine to give the Vandermonde determinant of the $U(N)$ matrix model, which is, modulo constant prefactors that we ignore, given by
\bel{\label{def delta}
  \Delta(\alpha) = \prod_{i > j} \sin^2 \frac{\alpha_i - \alpha_j}2.
}

A subtle and crucially important point arises when abelianizing the Polyakov loop: matrices $U(\b x)$ with equal eigenvalues represent degenerate points in the set of all Polyakov loop configurations. (We retain position dependence because the gauge must be fully fixed before we can meaningfully say that the integration of gauge fields fixes the $\alpha_i$'s to be constant.) We cannot perform the second gauge-fixing step outright, but instead we must break up the path integral over all $U(\b x)$'s into a sum over domains in which there are no eigenvalue overlaps anywhere on the sphere, and then abelianize in each domain separately. Remarkably, these domains are indexed by GNO charges. More precisely, it is possible to show that time-independent gauge transformations $\Lambda(\b x)$ needed to abelianize a generic nondegenerate $U(\b x)$ at all $\b x \in S^2$ need to be defined patch-wise on the two-sphere, just like in the Wu-Yang construction \cite{Jain:2013py}. These transformations $\Lambda(\b x)$ are then classified by the winding profile around the equator, and in turn this means that nondegenerate $U(\b x)$'s are classified by winding numbers of the gauge transformations needed to abelianize them. The winding numbers (a.k.a.~``flux sectors'') are given by $N$ numbers $m_i \in \Z$, one for each Cartan $H^i$ so that $e^{4\pi m_i H^i/\sqrt 2} = 1$. Performing a gauge transformation with nontrivial winding diagonalizes the holonomy $U(\b x)$ but simultaneously turns a smooth profile of the spatial gauge fields into a monopole configuration of form \eqref{def WY}. When evaluated on a configuration with a nontrivial winding of $U(\b x)$ and after the $\alpha(\b x)$'s are forced to be constant, the CS term gives the $\alpha$-dependent potential \cite{Aharony:2012ns, Jain:2013py, Kim:2009wb}
\bel{
  S\_{CS}(m, \alpha) = - ik m_i \alpha_i.
}

The integral over all Polyakov loops in the maximal torus gauge thus reduces to summing over flux sectors of spatial fields and integrating over all spatially constant configurations of eigenvalues $\alpha_i$. The sum over flux sectors gives
\bel{\label{fluxes}
  \sum_{m \in \Z^N} e^{-S\_{CS}(m, \alpha)} = \sum_{n \in \Z^N} \prod_{i = 1}^N \delta\left(\alpha_i - \frac{2\pi n_i}k\right).
}
In other words, this sum discretizes eigenvalues in units of $2\pi/k$, meaning that instead of working with an $N$-fold integral over all $\alpha \in [0,2\pi)$ we must work with a sum over $\alpha_i = 2\pi n_i/k$ for $n_i = 0, \ldots, k - 1$. In the planar limit, these eigenvalues are essentially continuous, although their discretization must be taken into account at large $\lambda$ \cite{Aharony:2012ns}.

It is important to emphasize that Polyakov loops associated to nontrivial flux sectors do \emph{not} correspond to the monopole backgrounds discussed in the previous section. We are talking about \emph{different} gauge-invariant field configurations here: disorder operators correspond to nontrivial profiles of the spatial components of the gauge field, while flux sectors label domains defined by eigenvalues of $U(\b x)$, the integral of the temporal component of the gauge field. It is only in the maximal torus gauge that some Polyakov loop configurations $U(\b x)$ manifest themselves as monopole backgrounds. In the next subsection we will explain how bona fide monopoles are to be included in the computation.

Finally, once the gauge is fixed and all other degrees of freedom have been integrated out, the partition function is given by the matrix model with discretized eigenvalues
\bel{\label{eq model}
  Z = \int [\d \alpha]\, \Delta(\alpha)\, e^{-4\pi T^2 v_0\left(e^{\alpha_i H^i\sqrt 2}\right)}.
}
This model is solved using the usual large $N$ methods, e.g.~\cite{Gross:1980he}, by replacing $\alpha_i$ with an eigenvalue density $\rho(\alpha)$ and carrying out a saddle-point calculation. This was done for all $\lambda$ in Refs.~\cite{Jain:2013py, Takimi:2013zca}. The four phases on Fig.~\ref{fig phases} are seen when $T^2 \sim N$ and they correspond to different qualitative behaviors of $\rho(\alpha)$: crossing the red critical line causes $\rho(\alpha)$ to lose support on the entire circle (the GWW transition), and crossing the blue line causes $\rho(\alpha)$ to cap off at $1/2\pi|\lambda|$, saturating the maximal possible value it can attain given the quantization of eigenvalues due to flux sectors (the DK transition).

\subsection{Monopole states}

Previous calculations of CS-matter partition functions using the above high-$T$ method did not sum over states with magnetic flux. The resulting partition function obeyed the dualities expected of the full partition function with monopoles included, however. This situation is similar to the calculation of the partition function for singlet vector models in \cite{Shenker:2011zf}. This partition function knew about the phase transition caused by baryons even though it was constructed just by counting single-trace operators, and in fact the $U(N)$ and $SU(N)$ results were the same even though the $U(N)$ theory did not even have baryons; this indicates that the contribution of baryon states to the partition function was negligible. The same is expected to happen for monopole states.

We now wish to take the sum over monopoles into account. We focus on those terms in this sum that correspond to states created by a single disorder operator with given GNO charges $\b q = (q_1, \ldots, q_N)$.  The calculation we need differs from the ones in earlier works in three ways:
\begin{enumerate}
  \item Some of the gauge symmetry needs to be spent on fixing the monopole background to the maximal torus form \eqref{def WY} at each spacetime point, as done in \cite{Kim:2009ia, Kim:2010ac}. For each set of $n$ equal GNO charges, a $U(n)$ factor contributes to the remaining gauge symmetry. Thus, for instance, if $q_1 = 1$ and all other charges are zero, the symmetry left to be fixed is $U(1) \times U(N - 1)$. If e.g.~$q_1 = q_2 = 1$ and all other charges are $q_i = 2$, the gauge symmetry is $U(2) \times U(N - 2)$. The Faddeev-Popov determinant from this gauge-fixing depends on the saddle-point configurations, not on the dynamical fields. The space of single-monopole backgrounds is discrete, and we focus on one specific background at a time; therefore the FP determinant cannot influence the physics of fluctuations around the saddle that we wish to study, and we may ignore it just as we ignore other constant prefactors. The reduced gauge symmetry influences the Vandermonde determinant $\Delta(\alpha)$ and the abelianization of the Polyakov loop. For instance, on backgrounds with all different GNO charges, the remaining gauge symmetry is $U(1)^N$, the Vandermonde is trivial, the contribution of the Polyakov loop to the action is just as if it was Abelian to begin with, and there are no flux sectors to sum over.
  \item High-energy modes will be insensitive to the flux background if the flux through a ``thermal cell'' of size $1/T^2$ is much less than unity. In other words, as long as each GNO charge satisfies $|q_i| \ll T^2$, the integral over high-energy modes proceeds just as in the case without monopoles. It is currently not known how to perform this integral at high GNO charges.
  \item The computation we wish to perform is reliable at high temperatures, but we are ultimately interested in extracting the Casimir energy at $T \rar 0$. We proceed by taking $T^2 = \zeta N$ and then working in the double scaling limit $N \rar \infty$, $\zeta \rar 0$. This is not as outlandish as it may seem: the theory has no phase transitions below a critical $T_c \sim \sqrt N$, so any computations in the regime $1 \ll T \ll T_c$ will be analytically connected to the $T \ll 1$ region. Indeed, Ref.~\cite{Jain:2013py} has shown that the $\zeta \rar 0$ limit of the free energy on a monopole-free background always reproduces the $T = 0$ result obtained by other means. In a monopole background the situation is only a bit more complicated. In the monopole-free sector, the dominant term of the partition function in the lowest-temperature phase is $e^{a T^4} \sim e^{N^2}$. In the case of interest for us, however, we are interested in the very small quantity $e^{- \Delta/T} \sim e^{-N^{\gamma}}$, and so in our double scaling limit we should expect to reproduce the $T = 0$ result up to corrections that have the same order in $N$. In other words, we expect to find the partition function of the form $e^{-\Delta/T} + c\, e^{-b N^\gamma}$ with $e^{-\Delta/T} \sim e^{-b N^\gamma}$. This is the form of the partition function we will recover, and we will be able to indicate which states contributed to the $e^{-b N^\gamma}$ correction.
\end{enumerate}

Computing the partition function in the single-monopole background is now straightforward, and we can borrow most of the technical results from earlier work. In particular, the first two points above show that the integral over high-energy modes does not depend on the monopole background as long as the GNO charges are much smaller than $T$. Thus, when setting up the matrix model with monopoles, in eq.~\eqref{def v(U)} we may use the result for the $\zeta \rar 0$ limit of the eigenvalue potential of CS-fermion theories obtained in eq.~(6.33) of Ref.~\cite{Jain:2013py},
\bel{
  v_0(U) = \frac1\pi \sum_{i = 1}^N \sum_{n = 1}^\infty \frac{(-1)^n}{n^3} \cos n\alpha_i.
}
The partition function in which this potential figures is a modification of \eqref{eq model} that takes into account the monopole background according to the three precepts above:
\bel{\label{eq general Z}
  Z_{\b q} = \int [\d \alpha]\, \Delta_{\b q}(\alpha)\, {\sum_{m_i}}' \, e^{ik\_{YM} \alpha_i q_i + i k \alpha_i m_i - 4\pi T^2 v_0(U)}.
}
In writing the above we arrange the GNO charges such that $q_1 \geq q_2 \geq \ldots \geq q_N$ with $q_i \in \Z$. For a general $\b q$, there are $p$ segments of $n_I$ equal charges (with $\sum_{I = 1}^p n_I = N$). The primed sum over flux sectors $m_i$ leaves out all the $m$'s that belong to a segment of length one, i.e.~all $m_i$ such that $q_i$ is not equal to any other GNO charge. The remaining fluxes are summed just like in eq.~\eqref{fluxes}. The Vandermonde is the one appropriate for the $\prod_{I = 1}^p U(n_I)$ group and is given by the product of determinants \eqref{def delta} for subsets $\alpha^{I}$ of eigenvalues  that are conjugate to the same GNO charge,
\bel{
  \Delta_{\b q}(\alpha) = \prod_{I = 1}^p \Delta(\alpha^{I}) = \prod_{I = 1}^p \prod_{1\leq i < j \leq n_I} \sin^2 \frac{\alpha^{I}_i - \alpha^{I}_{j}}2.
}

Eq.~\eqref{eq general Z} shows that monopoles with equal nonzero GNO charges are special: the sum over Polyakov loop windings $m_i$ will subsume part of the $q$-dependent potential $ik\_{YM} \alpha_i q_i $ in any sector with $n_I > 1$. The sum over $m_i$'s will still discretize the eigenvalues, and the remaining potential will depend on $\b q$ through the term $i N \alpha_i q_i$. It would be fascinating to analyze such monopole states, but the discretization of eigenvalues takes this problem outside the scope of the current work.

From now on we restrict ourselves to monopoles whose nonzero GNO charges are all different. For simplicity, let us consider $q_1 \neq 0$ and $q_i = 0$ for $i > 1$; the generalization to multiple nonzero charges will be trivial. The eigenvalue conjugate to $q_1$ now has special status, while the other eigenvalues $\~\alpha = (\alpha_2, \ldots, \alpha_N)$ can be treated just like in the previous subsection. In fact, the density of these $N - 1$ eigenvalues is unaffected by the dynamics of $\alpha_1$ at large $N$, and the density $\rho(\~\alpha)$ is determined by the very same saddle-point calculation used to find $\rho(\alpha)$ in the monopole-free case. The special eigenvalue $\alpha_1$ thus decouples, and the remaining eigenvalues are integrated over to give a multiplicative factor of $Z$, defined in eq.~\eqref{eq model}. The contribution of these eigenvalues to the partition function with monopoles represents the resummed single-trace fluctuations on top of monopole states, and in the low-$T$ limit it does not affect the Casimir energy $\Delta$. The remaining integral should be understood as an integral over all excitations of the matter dressing the monopole. In other words, the integral over the remaining eigenvalue represents the sum over all monopoles with the given GNO charge. This type of integral over a single eigenvalue and its relation to nonperturbative effects has, in fact, been observed in matrix models describing quantum gravity \cite{Neuberger:1980qh, David:1990sk, Shenker:1990uf}.

The partition function \eqref{eq general Z} is thus given by
\bel{\label{eq eval model}
  Z_{\b q} = Z \int_{0}^{2\pi} \d\alpha_1\ e^{ik\_{YM} q_1\alpha_1 - 4\pi T^2 v_0(\alpha_1)} = Z \int_{0}^{2\pi} \d\alpha_1\ e^{-N v\_{eff}(\alpha_1)},
}
with
\bel{
  v\_{eff}(\alpha_1) = - i \frac{1 - |\lambda|}\lambda q_1 \alpha_1 + 4\zeta \sum_{n = 1}^\infty (-1)^{n} \frac{\cos n\alpha_1}{n^3}.
}
The above integral can be solved using steepest descent. Before we proceed to find saddles of $v\_{eff}(\alpha_1)$, we note that CS-matter is invariant under the parity transformation $\alpha \mapsto -\alpha$, $k \mapsto -k$, and the above potential clearly respects this. From now on we take $\lambda > 0$; any saddle $\alpha_\star$ we find at a fixed positive $\lambda$ will correspond to a saddle $-\alpha_\star$ at $-\lambda$. For the same reason we also take $q_1 > 0$.

The saddle point equation is
\bel{
  i\frac {1 - \lambda}\lambda q_1 = 4\zeta \sum_{n = 1}^\infty (-1)^{n + 1} \frac{\sin n\alpha_1}{n^2}.
}
It is easy to find a complex saddle of the form $\alpha_1 = i y$, where $y$ is positive and large at small $\zeta$. To see that this is consistent and to find $y$, we substitute this ansatz and use $\sin n \alpha_1 = -\frac1{2i} e^{n y}$ in the saddle point equation, getting
\bel{
  \frac{1 - \lambda}\lambda q_1 = -2\zeta\, \trm{Li}_2 (-e^y) = \zeta \left(y^2 + O(y^0)\right),
}
and consequently
\bel{
  y = \sqrt{\frac{(1 - \lambda) q_1}{\lambda\zeta}} = \frac{\sqrt{q_1 k\_{YM}}}T,
}
which is indeed large as $\zeta \rar 0$. The value of the effective potential at this saddle point is
\bel{
  v\_{eff}(iy) = \frac23 \frac{(q_1k\_{YM})^{3/2}}{NT}.
}
We can now read off the Casimir energy $\Delta$ from $e^{-\Delta/T} = e^{-N v\_{eff}(iy)}$, giving the advertised result
\bel{
  \Delta = \frac23 (q_1k\_{YM})^{3/2}.
}

It can be checked that the steepest descent through this point in the complex plane is in the direction parallel to the real axis, and the initial contour can be deformed in the needed way without complications. There are no indications of instability around this saddle point. The contribution from the rest of the contour can be analytically shown to behave as $c\, T\, e^{-b T^2}$ for some coefficients $b$ and $c$.\footnote{It is amusing that in the na\"ive low-$T$ limit this contribution overwhelms the saddle point value and dominates the integral. This is confirmed by numerical evaluation of the integral \eqref{eq eval model}. There are no inconsistencies here because we are really taking the double-scaling limit in which $T$ is always large, as described in the third point at the beginning of this subsection. I thank Guy Gur-Ari and Ethan Dyer for numerous discussions and help with understanding the behavior of this integral away from the saddle point.} This term comes from parts of the contour far away from the saddle point, and we believe it can be understood as the partition function of highly excited monopole states --- the ones in which the dressing matter is not in its ground state but rather in states so energetic that their free energy eclipses the Casimir energy.

We now (very schematically) show how this contribution from excited monopoles can come about. Consider the integral over monopole states with dressing matter excited to an energy $E$ above the ground state. Their contribution to the partition function is
\bel{
  \int_0^\infty \rho(E) e^{-(\Delta + E)/T} \d E.
}
If $T$ is infinitesimal (meaning $T \ll 1$ on a spatial $S^2$ of unit radius), the above integral can be approximated just by $e^{-\Delta/T}$. However, as stressed in point 3) above, we work with $T = \sqrt{\zeta N} \gg 1$ and with $N \rar \infty$ taken before $\zeta \rar 0$. Thus, states with energies that differ by less than $T$ will not be suppressed relative to each other, and the cumulative effect of all these states can be enough to overwhelm the exponential suppression afforded by $\zeta \rar 0$. Consider high-energy excitations with $\rho(E) = e^{aE^{2/3}}$, the density of states appropriate for three-dimensional theories. A saddle point calculation shows that integrating over sufficiently high energy modes in the above integral gives
\bel{
  \delta Z = \int_\Lambda^\infty \rho(E) e^{-(\Delta + E)/T} \d E \propto e^{-b T^2},\quad b = \frac23 \left(\frac {(1 - \lambda)q}{\lambda\zeta}\right)^{3/2} - \frac{4a^3}{27}.
}
We take this as evidence that highly excited monopoles can, \emph{in principle}, sum to a number that cancels out the $e^{-\Delta/T}$ prefactor and gives the above formula for a small positive $b$. Only at $T \gg 1$ would it be possible to get $b$ of order one, so this is consistent with the expectation that our total result, $Z_{\b q} = Z\left(e^{-\Delta/T} + \delta Z\right)$, is analytically connected to $Z e^{-\Delta/T}$ that we would have gotten at $T \rar 0$ without the double-scaling limit, had we been able to carry out that calculation.  A careful study of monopole excitations would give us the correct density of states and allow us to check this conjecture.

%\bel{
%  \delta Z \sim T e^{-T^2},
%}
%exactly the kind of result the contour integral gave. Note that states with a single excited monopole (unlike the singlet excitations on top of a monopole) do not re-exponentiate, and each of them comes with its own prefactor of $Z$ stemming from singlet excitations on top of excited monopoles. This is all in perfect agreement with the total result we get,
%\bel{
%  Z_{\b q} = Z \left(e^{-\Delta/T} + c\, T e^{-b T^2} \right).
%}

If the monopole state has several different nonzero GNO charges, the same saddle-point equation is solved for each eigenvalue conjugate to these charges. The resulting Casimir energy can thus be recorded as the general answer for a monopole with few nonzero GNO charges:
\bel{\label{eq Delta general}
  \Delta = \frac23\left(\sum_i |q_i|^{3/2} \right) |k\_{YM}|^{3/2}.
}
This result stops being valid when there are so many nonzero GNO charges that the collective dynamics of their eigenvalues can no longer be neglected when studying the remaining eigenvalue density $\rho(\~\alpha)$. Moreover, if we had so many different charges, some of them would have charge comparable to $N$, thus invalidating the assumption that the magnetic flux in each thermal cell is negligible. Understanding high-flux monopole states remains an open problem.

\section{Outlook: new perspectives on bosonization}

We now possess a fair amount of information about the spectrum of operators at strong coupling in $U(N)$ CS-fermion theories. At low scaling dimensions we only have the tower of higher-spin $U(N)$ single-traces and multi-traces with $\lambda$-independent dimensions. Going to higher dimensions, we encounter the lightest disorder operators of dimension $\frac23 |k\_{YM}|^{3/2}$, and from there on we find more and more disorder operators with dimensions given by \eqref{eq Delta general}; there are also the monopoles with excited matter dressing whose dimensions we cannot calculate using the above method. (Disorder operators with several GNO charges equal to $1$ are not expected to have dimensions lower than the one with just one nonzero GNO charge $1$, but they might appear just above it in the spectrum; these have different topological charge, however, and we do not expect them to have nontrivial overlaps.) There are no baryon operators in $U(N)$ theories, so the disorder operators with $\b q = \left(\pm 1, 0, \ldots, 0\right)$ are the lowest-dimension nonperturbative operators.

What can these operators map to under the large $N$ CS-matter bosonization dualities? The dual to the stronly coupled CS-fermion theory of rank $N_F$ is the weakly coupled CS-Wilson-Fisher theory of rank $N_B$. (Recall that dual pairs of theories have ranks related by $N_B/|\lambda_B| = N_F/|\lambda_F| = |k|$.) All single-trace operators match across the duality, so the fermionic disorder operator cannot be dual to any combination of bosonic single-traces. Moreover, global symmetries must map across the duality, and the single-flux disorder operator is charged with a topological $U(1)_J$ symmetry \cite{Dyer:2013fja}. If the dual theory has gauge group $U(N_B)$, the only candidate in the weakly coupled bosonic theory is the bosonic disorder operator. If the dual theory is $SU(N_B)$, the only reasonable candidate is a baryon operator
\bel{
  B = \epsilon^{i_1\ldots i_{N_B}} \phi_{i_1} \del^{s_2} \phi_{i_2} \cdots \del^{s_{N_B}} \phi_{i_{N_B}}.
}
Based on level-rank duality of pure CS theory \cite{Jain:2013py}, we may expect that $U(N_F)$ maps to $U(N_B)$ while $SU(N_F)$ maps to $SU(N_B)$, meaning that disorder operators must map to disorder operators and baryons must map to baryons. However, in this speculative section we remain open to the possibility that $U(N_F)$ maps to $SU(N_B)$.

Can a bosonic baryon be dual to a fermionic disorder operator?  The powers $s_i$ of derivatives must be inserted so that the Levi-Civita symbol does not antisymmetrize the baryon to zero. Their positioning is similar to populating the Fermi sea with fermions in the disorder operator's matter dressing --- this is a pleasant feature but should not be taken to mean much, as the Fermi sea picture is not meaningful at strong coupling. More interestingly, though, this operator is charged under a baryon current $U(1)_B$, and in an $SU(N_B)$ theory this current is the natural (and only) candidate dual to the topological current $U(1)_J$ of disorder operators in $U(N_F)$ CS-fermion theories. A mapping between monopoles and baryons has been observed in some supersymmetric theories in three dimensions \cite{Aharony:2013dha, Aharony:2014uya}.

Even if some monopoles map to baryons under bosonization, the corresponding dimensions do not match up completely. The lowest bosonic baryon has dimension proportional to $N_B^{3/2}$ at weak coupling \cite{Shenker:2011zf}. The lowest fermionic disorder operator has dimension proportional to $k^{3/2}$. Their dimensions differ by a factor of $\lambda_B^{3/2}$. To make dimensions match without ruining the mapping of $U(1)$ charges, we must postulate that the fermionic disorder operator is bosonized to a bosonic baryon \emph{and} a sea of singlets needed to raise the dimension to the right number. In particular, a way that this can be done is to state that the lowest-dimension fermionic disorder operator maps to $M + 1$ baryons and $M$ antibaryons in a bosonic theory, with $M \sim \lambda_B^{-3/2} \sim k^{3/2} \gg 1$ at strong fermionic coupling. The $M$ baryon-antibaryon pairs can then be written as a combination of multi-traces. A similar multi-trace-dressing trick can be used to match up the dimensions of monopoles in the case of bosonic $U(N_B)$ duals to the $U(N_F)$ CS-fermion theory. However, this leaves open the question of what a single bosonic baryon is dual to.\footnote{I thank Guy Gur-Ari and Shiraz Minwalla for discussions concerning this conjectural baryon-monopole map. After the first version of this preprint was published, O.~Aharony understood the origin of this discrepancy and clarified the mapping of all these operators; see \cite{Aharony:2015mjs}.} The mismatch of baryon and monopole dimensions that comes from studying the lowest-dimension disorder operators may serve as evidence that bosonization really does map a $U(N_F)_k$ fermionic theory to a $U(N_B)_{-k}$ bosonic one, just as one would expect from pure CS level-rank duality.

It is particularly interesting to study these operator maps in light of the fact that the two phase transitions in Fig.~\ref{fig phases} are dual to each other \cite{Jain:2013py}. Based on general expectations of analyticity and the evidence from \cite{Shenker:2011zf}, at any $\lambda$ the GWW transition should be induced by baryon-antibaryon pairs that affect the multi-trace density of states. As all single-trace operators are dual to each other, the relations between them that cause the phase transitions should also be dual. The DK transition, being the dual of the GWW transition, should thus be induced by relations between single-traces, and in particular it is natural to assume that it arises due to either monopole-antimonopole or baryon-anitbaryon pairs affecting the density of states of single-traces. Monopole-antimonopole pairs can drive a phase transition via different ways to enact the OPE of disorder operators $\O(x)$ and $\bar \O(y)$ of opposite GNO charges; $\O\bar\O$ has zero topological charge so we can expect to find higher-spin currents in this OPE. In other words, a two-monopole state with zero topological charge may have nontrivial overlaps with multiple multi-trace states, and this may lead to relations between multi-traces.\footnote{Another argument for this goes as follows. In the planar limit there is no difference between $SU(N)$ and $U(N)$. The theories with these groups have the same phase structure, but only $U(N)$ has topologically charged monopoles, so once again we are led to think of topologically neutral configurations as the ones that drive the phase transition. In the same way, the baryonic phase transition exists in $U(N)$ theories in which baryons are not gauge-invariant operators.} It is thus very natural to conjecture that we can express a baryon-antibaryon pair $B\bar B$ in a fermionic theory as a pair of  nonperturbative bosonic operators. This is consistent with the picture of individual disorder operators mapping to individual baryons or monopoles, and it is now an interesting question of combinatorics (relegated to future work) to show whether the pairwise mapping of operators can be factored into mapping individual operators. Doing this combinatorics while respecting constraints set by the duality of phase transitions and the known dimensions of lowest-lying nonperturbative operators seems likely to reveal fascinating patterns in operator maps at finite $N$. If this is possible, we should also be able to conclusively decide whether the bosonization duality maps $U(N_F)$ to $U(N_B)$, to $SU(N_B)$, or to something else entirely.

Another question of interest is the bosonization of disorder operators with zero topological charge. Based on our calculations, at least some of these are also stable saddles of the path integral (e.g.~the dimension of the operator with $\b q = (q, -q, 0,\ldots,0)$ is $\frac43|qk\_{YM}|^{3/2}$). Thus they should not map to single- or multi-trace operators. A possible option is that they map to other disorder operators of zero topological charge.

Finally, a good check for the speculations given in this section would come from studying all the currently unknown quantum numbers of these nonperturbative operators. In particular, understanding bosonic disorder operators would be of great interest. The methods in this paper are not applicable to monopole states in CS-boson path integrals because scalar matter can condense even when the magnetic flux in a thermal cell is small. We view the generalization of the calculation in this paper to bosonic and supersymmetric theories as the most immediate and pressing topic for future work.

\section*{Acknowledgments}

It is a pleasure to thank Guy Gur-Ari, Ethan Dyer, and Jen Lin for collaborating on various stages of this project. Thanks are also due to Shamik Banerjee, Shiraz Minwalla, Steve Shenker, and Ran Yacoby for comments and useful conversations, and in particular to Ofer Aharony and Spenta Wadia for pointing out monopole regularization subtleties. The author is supported by a William R.~Hewlett Stanford Graduate Fellowship.

\end{document}